# Chirality in photonic systems

Dmitry Solnyshkov, Guillaume Malpuech

Institut Pascal, Photon-N2, Université Clermont Auvergne et CNRS, 4 avenue Blaise Pascal, 63178 Aubiere Cedex.

The optical modes of photonic structures are the so-called TE and TM modes which bring intrinsic spin-orbit coupling and chirality to these systems. This, combined with the unique flexibility of design of the photonic potential, and the possibility to mix photon states with excitonic resonances, sensitive to magnetic field and interactions, allows to achieve many phenomena, often analogous to other solid state systems. In this contribution, we review in a qualitative and comprehensive way several of these realizations, namely the optical spin Hall effect, the creation of spin currents protected by a non-trivial geometry, Berry curvature for photons, and the photonic/polaritonic topological insulator.

Quantum particles can be characterized by their mass, electric charge, spin, and energy. Artificial photon fluxes (energy fluxes) are realized since humanity mastered making fire. Classical charge currents were understood in the $19^{th}$ century and now form the basis of our modern world. Spin fluxes in photonic systems have also been created for quite a long time using polarizers and plates. In that case, there is no pure spin flux, because it cannot be dissociated from the photon/energy flux. In massive systems, the existence of the spin magnetic moment has been revealed by applying magnetic field in the Stern and Gerlach experiment [1], which was an example of a spin current associated with mass and energy current, created by using a magnetic field. In 1971, D'yakonov and Perel proposed a new scheme in which a pure spin current is created by the spin-orbit interaction in a semiconductor [2]. Electron charges are accelerated by an electric field along a wire, and scattered by impurities in transverse directions. The spin-orbit interaction results in the presence of an effective magnetic field acting on the electron spin only and with an orientation depending on the direction of propagation. So electrons which scatter in one direction align their spin with the corresponding field and electrons which scatter in the opposite direction align their spin parallel to an opposite field, which results in the creation of a transverse pure spin current. This effect has been called spin Hall effect, in analogy with the Hall effect, where a transverse charge current appears because of the Lorentz force acting on the charges due to a magnetic field. The question which then arises is whether such flux is possible without dissipation? The first example of dissipation-less transport came more than a century ago with the observation of superconductivity [3]. Superfluidity of helium was then reported 27 years later [4]. In both cases, a collective macroscopic flow is present, and the spectrum of single-particle excitations is deeply modified. The flow, even if it is not the ground state of the system, is energetically stable with respect to the creation of excitations. This suppresses elastic scattering by disorder because there are no final states available, and this suppresses the viscosity due to phonon scattering as well, the latter only leading to a certain finite depletion of the flow.

Another mechanism protecting a flow from elastic disorder scattering (but not from inelastic scattering with phonons) has been discovered more recently. It is based on the so-called "topological protection". Two systems are said to be topologically different (not homeomorphic) if one cannot pass from one to another by a continuous deformation (homeomorphism). For example, the topology of 2D closed surfaces is determined by the number of holes within this surface, a so-colled genus, which is a topological invariant, linked with the integral of the curvature – another topological invariant. A sphere and a torus are therefore topologically distinct objects. Another example of a topological invariant, which gives a better analogy for topological insulators discussed in this chapter, is the number of twists of a circular Möbius strip. The corresponding integral is the number of rotations of a vector normal to the surface of the strip (Figure 1a). A generalization of this invariant to the complex vectors leads to the Chern numbers, which we will discuss in the following.

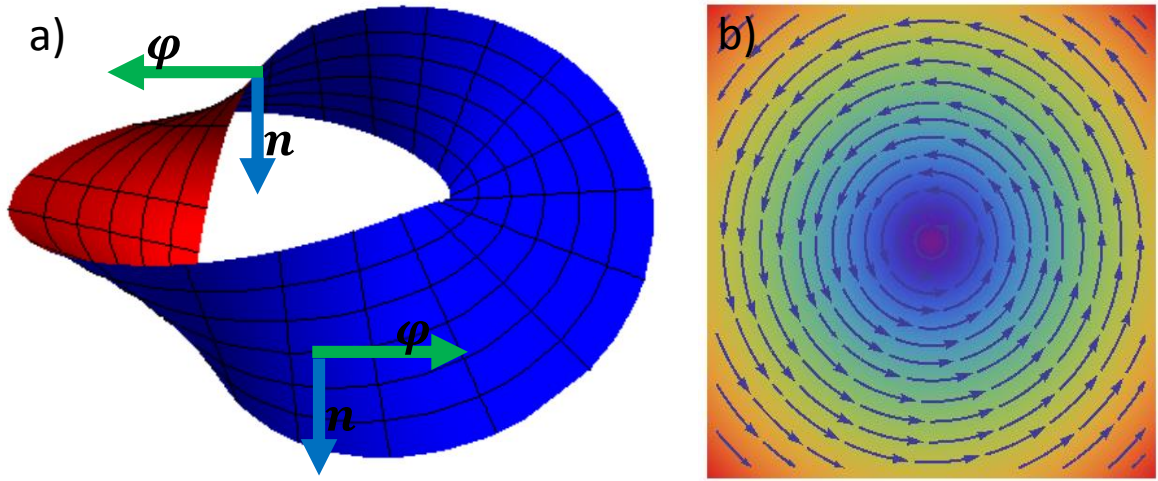

*Figure 1. Topological analogy of a Möbius strip (a) and a vortex (b): in both cases there is a 360 degrees rotation of the normal angle **n** around the spherical unit vector **φ** (a) or the phase angle (b).*

In Physics, well-known topologically non-trivial objects are vortices in interacting quantum fluids. These vortices are characterized by a quantized circulation, because they are described by a complex wavefunction. More importantly, the core of the vortex must have zero density, and it cannot be smoothly removed to reach a constant solution. A quantum vortex is therefore associated with a persistent rotating flow of particles, whose rotation cannot be dissipated because of the topological protection. The velocity of the flow, similar to the direction of the normal of the full-twist Möbius strip, makes a full turn when going around the center (Fig. 1b). It has been understood at the end of the seventies that the concept of topological classification also applies to the states of energy bands in periodic systems such as solids. Berry first [5] has shown that the evolution of a wavefunction in a parameter space leads to the accumulation of an extra phase in the wavefunction, the Berry phase, which in fact was already known in optics as the Pancharatnam phase [6]. It was then understood that the Berry phase accumulation over all the states of a complete energy band is quantized, exactly as a geometrical curvature of a closed body, and relies to a topological invariant (the Chern number) defined in the forties by Chern for complex vectors [7]. These discoveries were triggered by key experimental results, namely the demonstration of the integer Quantum Hall effect [8]. It was understood that the quantized Hall conductance of each Landau band is nothing but the Chern number of the band.

It took some more time to understand that the presence of these topologically non-trivial bands imposes on the interface with a topologically different system, such as vacuum, the presence of chiral edge modes [9] having their dispersion joining two successive Landau bands. The qualitative argument explaining the existence of these edge modes is that passing from a topological to a trivial insulator imposes closing the gap on interface. The chiral character of the modes is due to the vorticity imposed to the system by the magnetic field. The surface state must bridge two successive Landau levels with different vorticity. Electrons propagating in these states can neither backscatter nor resonantly scatter to the bulk, because of the absence of the corresponding states. The understanding of all these concepts brought into Physics the notion of a topological insulator: an insulating system with non-trivial topology of the bands, necessarily accompanied by chiral edge states. Figure 2 helps to understand this by presenting a sketch of a pair of topologically nontrivial bands (e.g. in a massive Dirac equation [10]) with the edge states appearing at the borders of a ribbon-shaped sample. These edge states arise from the allowed band states having the same pseudospin, and therefore necessarily possess a certain group velocity. If a usual conductor or a semiconductor can be represented by an hourglass filled with sand, a topological insulator is a twisted hourglass, where a grain of sand moving vertically in the neck necessarily moves also in the horizontal plane.

In the following, several key results have been obtained. Haldane proposed theoretically in 1988 [11] the anomalous quantum Hall effect in a graphene layer placed in a spatially alternating magnetic flux with a zero average. He showed the existence of one way topologically protected edge modes without

Landau levels, demonstrating that the concept of a topologically non-trivial band is not uniquely related to Landau levels. A very strong breakthrough came with the discovery of the Quantum Spin Hall effect ($Z_2$ topological insulator, as explained below), where a specific spin-orbit coupling, an effective momentum-dependent magnetic field, creates an opposite vorticity for each of the spin components [12]. In total, there is no vorticity in the system, but for each spin component there is one, and the *spin* Chern numbers are non-zero. This leads to the formation of a topologically protected spin current on the surface of such topological insulator. Currently, the terminology is not completely established yet: the $Z_2$ topological insulators (where the sum of the spin Chern numbers makes up an integer topological invariant being zero) are often called simply "topological insulators", whereas the Z topological insulators (with non-zero integer band Chern numbers) are also called "Chern insulators". Besides, one should note that the concept of the $Z_2$ topological insulator is based on the fact that the spin of an electron is a well-defined conserved quantum number, so that a spin-up electron cannot reverse its spin during a scattering process, unless there is a magnetic impurity.

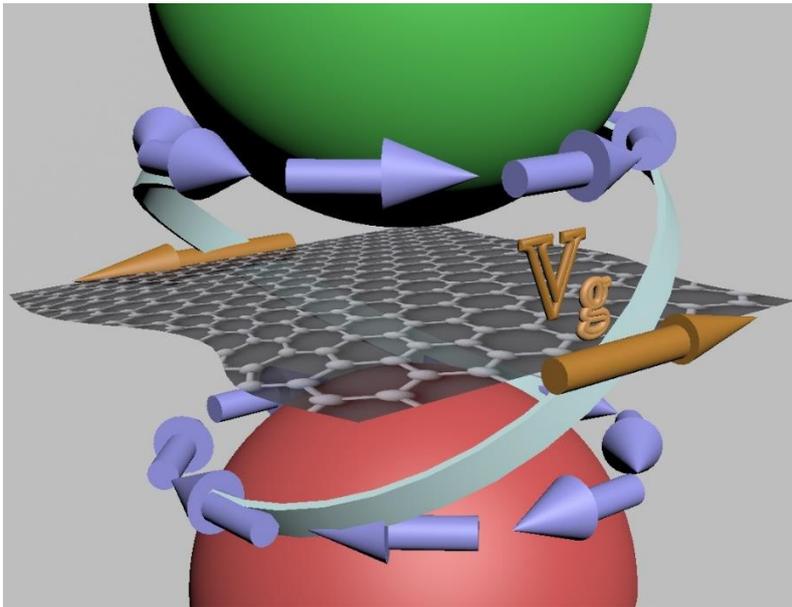

*Figure 2. Vorticity and edge states in topological insulators. The red and green paraboloids are valence and conduction bands, blue arrows show the pseudospin orientation, light blue line is the edge state connecting two identical pseudospins, and brown arrow shows the group velocity. The ribbon in the middle is the finite-size sample.*

Finally, in 2008, Haldane and Raghu [13] made the concept of topology enter the field of photonics. They showed that the peculiar structure of the TE-TM modes in a photonic system allows the existence of topologically non-trivial photonic bands and predicted a dissipation-less transport of photons on the edge of a photonic crystal, a proposal which has been demonstrated experimentally soon after [14] and extensively studied since then [15]. Fractional Quantum Hall states in photonics have also been proposed in 2008 [16]. As in electronic topological insulators, both normal and anomalous quantum Hall effects can be used in topological photonics to achieve the non-trivial insulating regime [17]. Since photons are not charged and therefore are not affected by real magnetic fields, for the normal quantum Hall effect one has to create artificial magnetic (gauge) fields in photonic structures [18,19,20,21]. While initially the concept of topological insulators was mostly applied to the periodic structures with non-trivial topology of the Bloch bands, later it was extended to more general band structures, which appear e.g. in chiral hyperbolic metamaterials [22]. Indeed, a Dirac point appears naturally in the dispersion of a non-periodic medium at $k=0$ in presence of the Rashba spin-orbit coupling (because the latter is linear in wavevector), which can emerge in the chiral media [23]. A topological gap can then be opened at this Dirac point. Spin-orbit coupling exists even for light waves in the vacuum: since such waves can only be transverse, their polarization is linked with the propagation direction. This coupling can also be employed to obtain the Quantum spin Hall effect for photons, if the corresponding pseudospin is conveniently defined [24].

In this work, we are going to review some of the basic concepts of topological photonics, essential for the works cited above. We first remind the concept of effective magnetic fields acting on a pseudospin. We then discuss a few examples: the case of the Rashba field acting on the electron spin in semiconductors, the case of the monopolar field acting on the lattice pseudospin in graphene, and the effective TE-TM field acting on the photon pseudospin. We will then remind a few consequences of the TE-TM field, such as the optical spin Hall effect [25], the creation of a non-dissipative spin current in a geometrically non-trivial system (ring) [26]. We then remind that the combination of TE-TM and Zeeman effective field induces a non-zero Berry phase when a photon is going along a curved trajectory both in real and reciprocal space [27]. We will then consider two photonic topological insulators based on this non-zero Berry curvature and on the use of a proper periodic potential. The first case is the one of Haldane-Raghu in a photonic crystal. The second one [28,29] is based on the use of a lattice of coupled 0D photonic resonators.

**I Concept of effective magnetic field**

The electron spinor (spin ½) is a two-dimensional complex vector which evolves according to a $2\times 2$ Hamiltonian. The coupling of the electron dipole moment with a magnetic field is described by a scalar product between the magnetic field and a vector of Pauli matrices. It gives rise to a well-known dynamics: Larmor precession of the spin around the field. But in fact, any $2\times 2$ Hermitian matrix can be decomposed into a linear combination of Pauli matrices and an identity matrix:

$$\hat{H} = \begin{pmatrix} H_0 + \dfrac{\Omega_z}{2} & \dfrac{\Omega_x - i\Omega_y}{2} \\ \dfrac{\Omega_x + i\Omega_y}{2} & H_0 - \dfrac{\Omega_z}{2} \end{pmatrix} = H_0 \mathbf{I} + \frac{1}{2}\vec{\Omega}_{Total} \cdot \vec{\boldsymbol{\sigma}} \qquad (1)$$

The dynamics of any two level system can therefore be physically described as an interaction between an effective magnetic field $\vec{\Omega}_{total}$ and a 3-component pseudospin vector generated by the spin operators $\sigma_x, \sigma_y, \sigma_z$ and defined via the wavefunction components as $\vec{S} = \left(\text{Re}(\psi_+\psi_-^*), \text{Im}(\psi_+^*\psi_-), (|\psi_+|^2 - |\psi_-|^2)/2\right)^T$. This formalism is extremely widely used in order to describe light-matter coupling in atoms or quantum dots, the electronic band structure of graphene, spin-orbit interaction for electrons in solids, and the polarization states of light in photonic systems, such as waveguides, cavities, or photonic crystals. In electronic systems, the unit sphere of the electron spin is typically called a Bloch sphere, whereas in photonic systems the dedicated name is Poincaré sphere. Let us now consider a few examples.

**I-a The case of Electrons in 2D semiconductors (Rashba field)**

Spin-orbit coupling means that the spin degree of freedom is coupled with the momentum of the particle (in general, both its magnitude and orientation). The Rashba and Dresselhaus spin-orbit couplings appear in semiconductors under symmetry breaking. Structural inversion asymmetry, responsible for the Rashba field, emerges, for example, in quantum wells under electric field applied in the growth direction, while the bulk inversion asymmetry, necessary for the Dresselhaus field, is present in structures that lack inversion symmetry, e.g. zinc-blende lattice with two different atoms (GaAs). Qualitatively, the Rashba effect [30] can be understood along the same lines as the spin-orbit coupling in atoms and the spin Hall effect: the relativistic field transformation converts the applied electric field $E_z$ into a magnetic field in the frame of an electron propagating with the velocity $\mathbf{v}$ in the XY plane:

$$\boldsymbol{\Omega} = -(\mathbf{v} \times \mathbf{E})/c^2 \qquad (2)$$

where $c$ is the speed of light in the vacuum. This magnetic field is then coupled to the spin of the electron via the usual term $-\mathbf{\Omega}\boldsymbol{\sigma}$ which reads:

$$H_{SO} = \frac{g\mu_B}{2c^2}(\mathbf{v}\times\mathbf{E})\cdot\boldsymbol{\sigma} = \frac{g\mu_B E_z}{2mc^2}(\boldsymbol{\sigma}\times\mathbf{p})\cdot\mathbf{u}_z \qquad (3)$$

where $g$ is the electron g-factor, $\mu_B$ is the Bohr's magneton, $m$ is the electron mass, $\mathbf{u}_z$ is the unit vector in the Z direction. The exact expression for the prefactor depends on the material parameters. The texture of the Dresselhaus field [31] is different from that of the Rashba field, and both fields together, which can be present e.g. in GaAs QWs under applied field, are usually included in a Hamiltonian as follows:

$$H \approx \frac{\hbar^2 k^2}{2m} + \alpha\left(k_x\sigma_y - k_y\sigma_x\right) + \beta\left(k_x\sigma_x - k_y\sigma_y\right) \qquad (4)$$

where $\alpha, \beta$ are the coefficients of the Rashba and Dresselhaus fields respectively. The textures of these fields are shown in Fig. 3, together with the texture of the monopolar field.

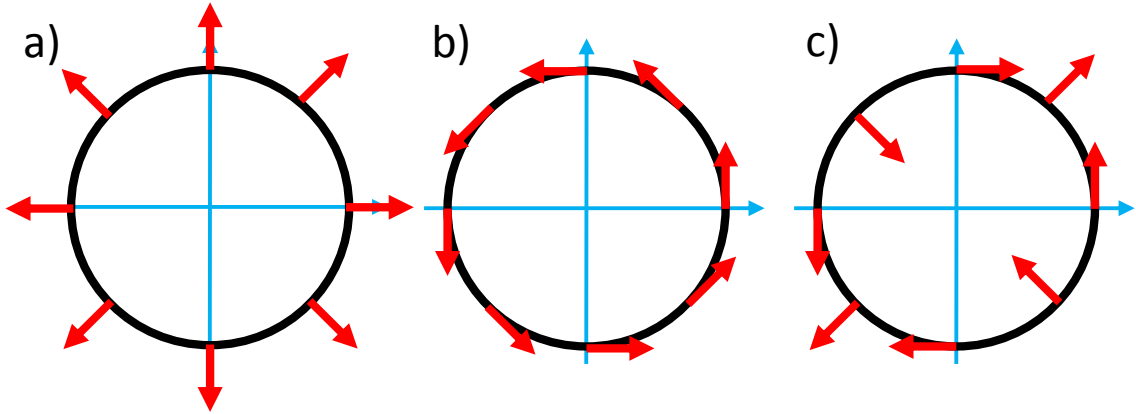

*Figure 3. Textures of several possible single-winding ($2\pi$ rotation of the field for a $2\pi$ rotation of the wavevector) effective fields: a) monopolar; b) Rashba; c) Dresselhaus.*

### I-b The case of Graphene

The 2D graphene lattice (Fig. 4a) contains two different types of inequivalent atoms A and B. An atom A is surrounded only by atoms B and an atom B is surrounded only by atoms A.

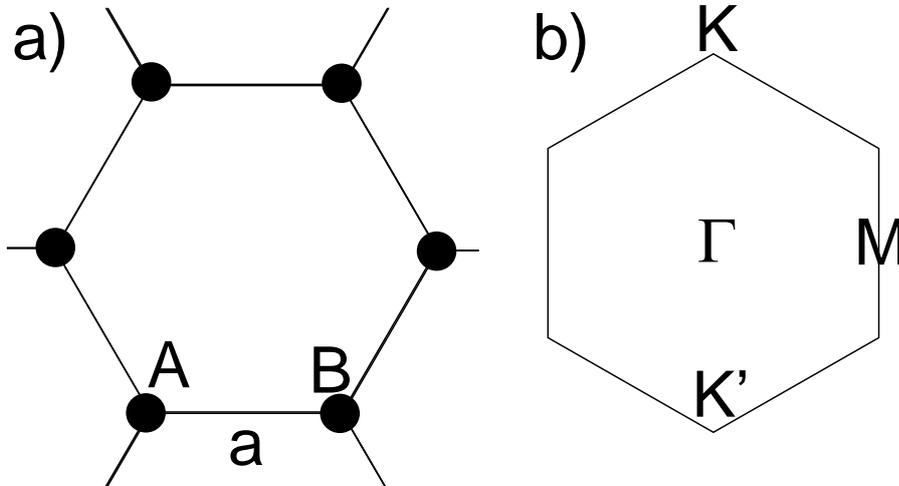

*Figure 4. a) Scheme of the honeycomb lattice of graphene in real space; b) the first Brillouin zone of the graphene lattice in the reciprocal space, showing the most important high symmetry points ($\Gamma$ is*

*the Brillouin zone center with parabolic dispersion, $K, K'$ are the zone corners with linear conical dispersion – the so-called Dirac points, and $M$ is a saddle point at the zone edge).*

Therefore, on the basis of A and B orbitals, in a tight binding approach and neglecting second neighbor interaction, the Hamiltonian takes the following form:

$$H_{graphene} = \begin{pmatrix} 0 & Jf_k \\ Jf_k^+ & 0 \end{pmatrix} = J\left(\text{Re}(f_k)\sigma_x - \text{Im}(f_k)\sigma_y\right) \tag{5}$$

where $J$ is the nearest-neighbor tunneling coefficient in the tight-binding model, $f_k = \sum_{j=0}^{2} \exp\left(ia\vec{k}.\left(\cos\left(\frac{2j\pi}{3}\right)\vec{u}_x + \sin\left(\frac{2j\pi}{3}\right)\vec{u}_y\right)\right)$, $a$ is the distance between the nearest neighbors (see Fig. 4a), $\vec{k}$ is the wavevector, $\vec{u}_{x,y}$ are the unity vectors along the X and Y axes, correspondingly. In this system, the pseudospin is by no means related to the real spin of the particles: it defines the correlations between A and B atoms. The eigenstates of the system are found through the alignment and anti-alignment of the pseudospin with a $\vec{k}$-dependent effective magnetic field $\vec{\Omega}_{graphene} = J\left(\text{Re}(f_k), -\text{Im}(f_k), 0\right)^T$. The states with negative energy have their pseudospin aligned with this field, and their energy lies below the Dirac points (valence band). The ones with their pseudospin anti-aligned have a positive energy and are above the Dirac points. The effective field cancels at the Dirac points and depends linearly on $k$ around it: $\Omega_{graphene} \equiv \left(\tau_z k_x, k_y, 0\right)^T$, which provides the famous linear Dirac dispersion for electrons, with $\tau_z = \pm 1$ being the valley index (+1 for K, -1 for K', see Fig. 4b). One can observe that the field is monopolar for $\tau_z = +1$ and Dresselhaus-like for $\tau_z = -1$, and therefore has an opposite winding at $K$ and $K'$: the rotation directions of the vortices in these points are opposite. If we use the analogy with the Möbius strip, there are two twists in the opposite directions, and the resulting system is not twisted at all. A breaking of the spatial inversion symmetry (e.g. making A and B atoms different) can open a gap in such structure, but the Berry curvatures, discussed below, although being nonzero for the two valleys, have opposite signs, the Chern numbers of the two bands are zero in that case, and the gap is trivial, despite the local vorticity of the wavefunction.

**I-c The case of Photons in photonic structures** [32]
The linear polarization of the transverse waves of the electromagnetic field comes from the superposition of the two counter-rotating circular components. The relative phase between them determines the direction of oscillation of the electric field in the plane perpendicular to the direction of propagation. In the vacuum, all allowed polarizations are degenerate, which is not the case in confined photonic systems, such as microcavities or waveguides. In these systems, in the absence of magnetic activity, the eigenmodes are the TE and TM, and they are split in energy. This splitting comes from the difference of the phase shifts of the two polarizations experienced during internal reflection ($\Delta\varphi_{TM} = \Delta\varphi_{TE} n_1^2 / n_2^2$). It is therefore a very fundamental effect, appearing at all interfaces between media with different refraction coefficients. In microcavities, it can be further controlled by the structure asymmetry [33].

On a circular polarization basis, the Hamiltonian describing this splitting for a 2-component photonic spinor reads:

$$H = \begin{pmatrix} H_0(\vec{k}) & \Omega_{TETM}(k)e^{-2i\varphi} \\ \Omega_{TETM}(k)e^{2i\varphi} & H_0(\vec{k}) \end{pmatrix} = H_0(\vec{k})I + \vec{\Omega}_{TETM}.\vec{\sigma} \tag{6}$$

$\vec{k}$ is a 2D wavevector, with $\varphi$ being the polar angle. $\Omega_{TETM}(k)$ is the magnitude of the energy splitting between the TE and TM modes. $\vec{\sigma}$ is a vector of Pauli matrices, $\vec{\Omega}_{TETM}(\vec{k}) = \Omega_{TETM}(k)(\cos 2\varphi \quad \sin 2\varphi \quad 0)^T$ is the TE-TM effective magnetic field (the Z component is zero, because the field is in-plane). The general texture of the field in reciprocal space around a point of degeneracy ($k=0$ in a planar cavity) is the one of a dipolar field (figure 5a). At a fixed amplitude of the wavevector, the field has a winding number 2 (figure 5b).

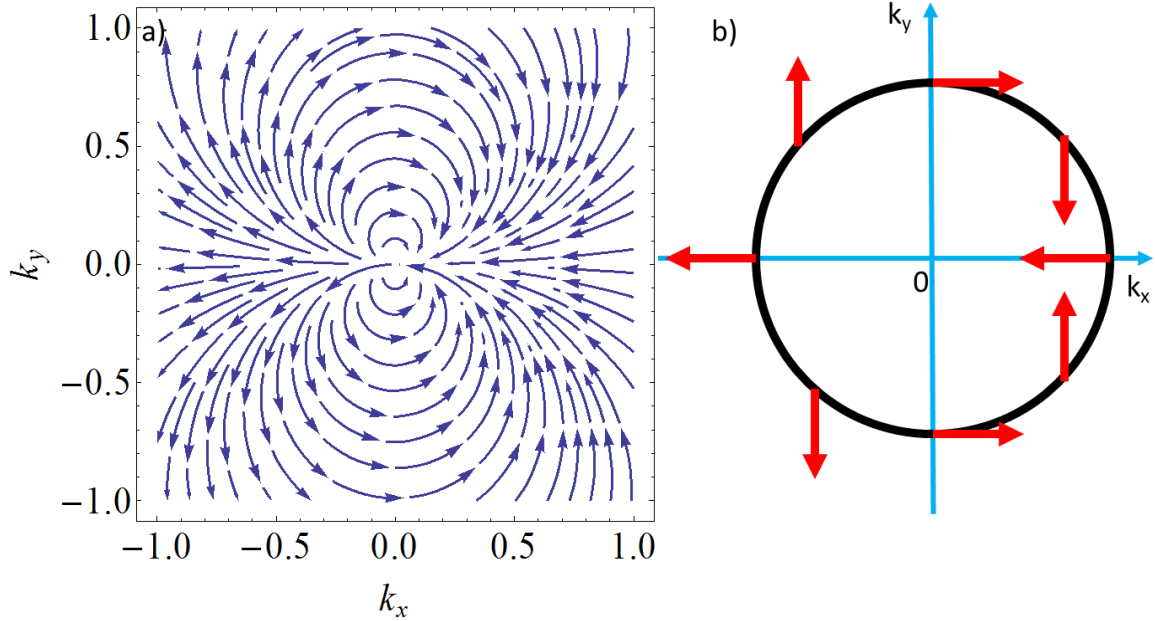

*Figure 5. a) The texture of the TE-TM field in the reciprocal space. b) The orientation of the TE-TM field for a fixed wavevector magnitude (isoenergetic circle), demonstrating a double winding.*

The photonic wavefunction of a TE mode having a pseudospin aligned with the effective field is given by

$$\vec{\psi}(\vec{k}) = \psi(k)\begin{pmatrix} e^{-i\varphi} \\ e^{i\varphi} \end{pmatrix} \qquad (7)$$

One can see that for a rotation in the reciprocal space the relative phase changes continuously as $2\varphi$. This is the required condition for the mode to remain transverse-polarized (TE). If one compares $\varphi=0$ and $\varphi=\pi$, the relative phase between the two circular components changes by $2\pi$ so that the polarization at these two points is the same. However, the total wavefunction has changed sign. A similar behavior arises for electrons in a Rashba field if they make a full rotation in reciprocal space. This difference evidences the nature of the two types of particles: fermionic for electrons and bosonic for photons. If we turn to the analogy of a Möbius strip, fermions correspond to a half-integer-twisted strip, which does not return to the original configuration after a full $2\pi$ rotation. A boson is similar to an integer-twisted strip, which is self-similar after a $2\pi$ rotation.

## II Consequences of the Chirality induced by the TE-TM effective field
### II-a Optical Spin Hall effect [25]

The structure of the effective field acting on photons can be evidenced through the optical spin Hall effect. If a radial flow of linearly polarized isoenergetic photons is created by a laser with a finite spot size (Fig. 6a), the polarization precession about the wavevector-dependent effective field gives rise to a circularly polarized emission pattern in real space, with most of the light emitted during the first pseudospin oscillation because of the finite photon lifetime. Four spin domains therefore appear in real space (Fig. 6b,c). Photons propagating in opposite directions have the same spin. Opposite spins are achieved for perpendicular propagation directions. This is different from the electronic spin Hall effect, where electrons propagating in opposite directions have opposite spins.

Another interesting scheme takes place when the elastic circle (isoenergetic line in the reciprocal space, accessible via the elastic scattering mechanism) is excited by a circularly polarized light (pseudospin along z) which then precesses about the effective field [34]. The pseudospin gains a $\varphi$-dependent in-plane projection perpendicular to the $\varphi$-dependent effective field. This leads to the appearance of a relative phase $2\varphi$ between the two circular components. After a half-turn of the pseudospin, the wavefunction becomes: $\psi = \begin{pmatrix} 0 & e^{-2i\varphi} \end{pmatrix}^T$ which is a vortex of winding number -2 for the $\sigma^-$ component.

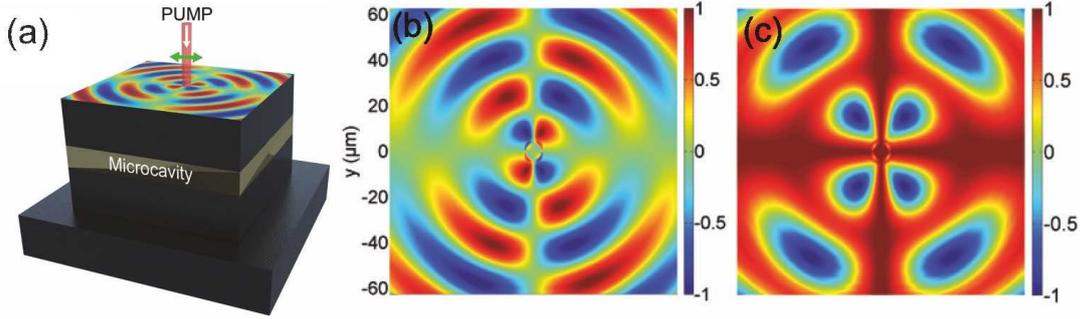

*Figure 6. The optical spin Hall effect with linear polarized pumping: a) scheme of the experiment; b) circular polarization degree (color) as a function of 2 spatial coordinates, c) linear polarization degree (color) as a function of 2 spatial coordinates.*

Another configuration, where the spin-orbit coupling affects the trajectory of photons, corresponds to that of a finite-size polarized beam propagating in a certain direction through a disordered medium and being deviated due to the TE-TM or a similar effective field arising from the spin-orbit coupling of light [35,36].

**II-b Topologically protected spin current in ring-like geometries [26]**
Let us now reduce the dimensionality, going to a 1D case with periodic boundary conditions (Fig. 7a). Without polarization splitting, the isoenergetic subset is composed of four states:

$$\psi_+ = \begin{pmatrix} e^{\pm i|k_+|\frac{2\pi}{L}x} \\ 0 \end{pmatrix}, \psi_- = \begin{pmatrix} 0 \\ e^{\pm i|k_-|\frac{2\pi}{L}x} \end{pmatrix} \qquad (8)$$

with $|k_+| = |k_-|$. In the presence of the TE-TM splitting, which, in this case, is just a constant magnetic field in the X direction, the eigenstates should have their pseudospin aligned or anti-aligned with the field and become:

$$\psi_x = \frac{1}{\sqrt{2}} \begin{pmatrix} e^{\pm i|k_+|\frac{2\pi}{L}x} \\ e^{\pm i|k_-|\frac{2\pi}{L}x} \end{pmatrix}, \psi_y = \frac{1}{\sqrt{2}} \begin{pmatrix} e^{\pm i|k_+|\frac{2\pi}{L}x} \\ -e^{\pm i|k_-|\frac{2\pi}{L}x} \end{pmatrix} \qquad (9)$$

where again $|k_+| = |k_-|$. These two states are split by the polarization splitting. They do not carry any spin current, because both $\sigma^+$ and $\sigma^-$ have the same wave number. There is, for each polarization, a doublet of states with two opposite directions of propagation. In case of an infinitely long wire, the X and Y-polarized states form two bands, and the isoenergetic space is formed by four states, two counter propagating X-polarized states with a given amplitude of the wave number and two counter propagating Y-polarized states with another magnitude of the wave number.

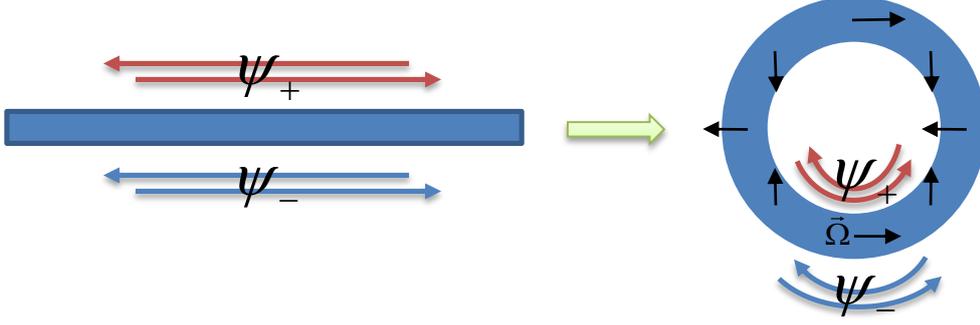

*Figure 7. 1D wire cavity and a ring cavity with the TE-TM splitting*

Let us now bend the wire in order to make a ring (Fig. 7b). Without TE-TM splitting, the isoenergetic space is again formed by four states. For each circular polarization, the angular momentum, which replaces the wave number, can take two opposite values, with the wavefunctions being

$$\psi_{l_+}^+ = \begin{pmatrix} e^{\pm i|l_+|\varphi} \\ 0 \end{pmatrix}, \psi_{l_-}^- = \begin{pmatrix} 0 \\ e^{\pm i|l_-|\varphi} \end{pmatrix} \qquad (10)$$

Let us now include the TE-TM splitting, keeping it much smaller than the splitting between the quantized states. We insert into the Hamiltonian the scalar product of the pseudospin of a state and an effective field corresponding to the polarization locally parallel to the wire axis:

$$\Omega(\varphi) = \Omega_0 \begin{pmatrix} \cos 2\varphi \\ \sin 2\varphi \end{pmatrix} \qquad (11)$$

It is easy to see that the product is nonzero if $l_+ = l_- + 2$. Let us now assume a small ring with the kinetic energy much larger than the polarization splitting, and consider the quadruplet $|l|=1$. One can see that the state $l_+ = 1$ gets coupled to $l_- = -1$. On the other hand, $l_+ = -1$ and $l_- = 1$ remain uncoupled with the states of this subset. The four new eigenstates read:

$$\psi_{TE} = \frac{1}{\sqrt{2}} \begin{pmatrix} e^{i\varphi} \\ e^{-i\varphi} \end{pmatrix}, \psi_{TM} = \frac{1}{\sqrt{2}} \begin{pmatrix} e^{i\varphi} \\ -e^{-i\varphi} \end{pmatrix}, \psi_{-1}^+ = \begin{pmatrix} e^{-i\varphi} \\ 0 \end{pmatrix}, \psi_1^- = \begin{pmatrix} 0 \\ e^{i\varphi} \end{pmatrix} \qquad (12)$$

The TE and TM modes are energetically split between themselves and from the two other states by the polarization splitting.

In this geometry, it is not possible to separate anymore the motional and spin degrees of freedom, contrary to the case of the linear wire, where the two states propagating in one direction and the two ones propagating in the other direction formed two decoupled spinors interacting with an effective magnetic field. One can also observe that each of these four states (Eq. (12)) carries a spin current which cannot be elastically backscattered, because the corresponding states are split-off in energy from all others. Regarding the two uncoupled states (which are degenerate and so can form any linear combination), they cannot be a source of spin current inversion in the TE and TM states as well, because the spin is conserved during the scattering on a potential.

An example of such spin current, protected by the nontrivial topology of the system itself, that is, the presence of a hole leading to a finite energy splitting between the eigenstates carrying opposite spin currents, has been recently explored experimentally and theoretically in the work studying a polariton benzene molecule [26]. Polaritons appear in microcavities in the regime of strong coupling of excitons and photons. Thanks to their excitonic fraction they can efficiently interact with the medium and with each other, which provides their thermalization. In this system, condensation occurred spontaneously on a single state carrying a nonzero spin current. A calculated image of the condensate density, together with the corresponding spin current distribution and the linear polarization orientation is shown in Fig. 8.

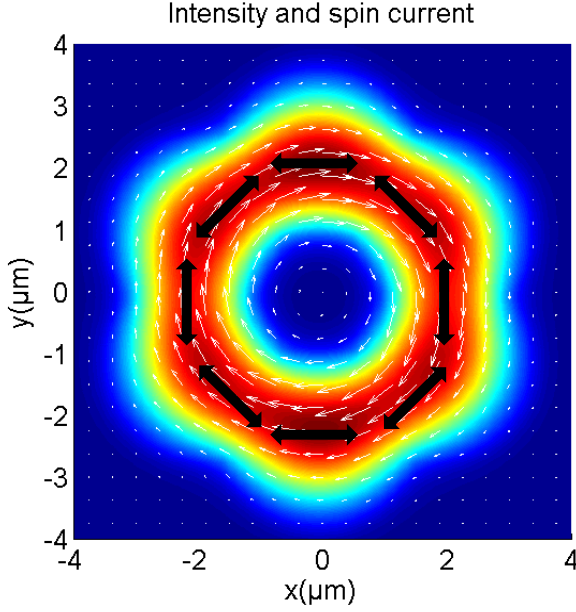

Figure 8. Density (color), spin current (white arrows), and linear polarization (black arrows) of a condensate formed in the lowest substate of the $|l|=1$ quadruplet in a ring-like polaritonic structure (benzene molecule [26]).

**II-c Combination of TE-TM and Zeeman field: Berry phase for photons [27]**
When an external magnetic field is applied perpendicularly to the plane of a 2D structure, a weak Zeeman-like effect is taking place for purely photonic systems. It is much larger in strongly coupled exciton-polariton systems, because of the excitonic Zeeman splitting. The total effective field acting on polaritons reads $\Omega = (\Omega_{TETM}(k)\cos 2\varphi, \Omega_{TETM}(k)\sin 2\varphi, \Omega_z)^T$. The circular polarization degree of the photon emission $\rho_c = (I_+ - I_-)/(I_+ + I_-)$, where $I_\pm$ is the intensity of the circular components, corresponding to the polarization pseudospin aligned with the field, reads $\rho_c = \Omega_z / \sqrt{\Omega_z^2 + \Omega_{TE-TM}^2}$, and the wavefunction spinor is given by:

$$\psi = \frac{1}{\sqrt{2}}\begin{pmatrix} \sqrt{1+\rho_c}\, e^{-2i\varphi} \\ \sqrt{1-\rho_c} \end{pmatrix} \quad (13)$$

We can calculate the Berry phase for a closed trajectory either in direct or in reciprocal space: both involve the rotation of the polar angle $\varphi$ determining the propagation direction. Using the standard definition,

$$\gamma = i\int \langle \psi | \nabla_R \psi \rangle dR \quad (14)$$

we obtain

$$\gamma = i\int_0^\varphi \left\langle \psi(\varphi') \left| \frac{\partial}{\partial \varphi'} \right| \psi(\varphi') \right\rangle d\varphi' = (1+\rho_c)\varphi \quad (15)$$

We see that for zero circular polarization degree Berry phase is simply given by the rotation angle in the real space: it is $2\pi$ for full rotation. For a nonzero magnetic field, Berry phase for a full circle or between two half-circles starts to be different, and it can therefore be measured by interferometry [27]. This Berry phase of $2\pi$ can also be included in the spinor, which then becomes completely symmetric:

$$\psi = \frac{1}{\sqrt{2}}\begin{pmatrix} \sqrt{1+\rho_c}\, e^{-i\varphi} \\ \sqrt{1-\rho_c}\, e^{+i\varphi} \end{pmatrix} \quad (16)$$

Using the Stokes' theorem, Berry phase for a closed contour can also be written as an integral of the gauge-invariant Berry curvature, which determines the topology of the eigenstates, exactly as the Gauss curvature determines the topology of real objects. From the point of view of the Stokes' theorem, Berry connection plays the role of a vector potential $\mathbf{A} = i\langle\psi|\nabla_R\psi\rangle$, which is gauge-dependent, while the Berry curvature is equivalent to a magnetic field $\mathbf{B} = \nabla_R \times \langle\psi|\nabla_R\psi\rangle$, which is gauge-invariant. The parameter space $\rho_c, \varphi$ can be replaced by two angles on a sphere using $\rho_c = -\cos\theta$. In this parameter space with spherical coordinates, the radial component of the Berry curvature (which is the only one different from zero) is given by:

$$B_r = \frac{1}{\sin\theta}\frac{\partial}{\partial\theta}\left(\sin\theta\left\langle\begin{array}{c}\sqrt{\frac{1-\cos\theta}{2}}e^{-2i\varphi}\\\sqrt{\frac{1+\cos\theta}{2}}\end{array}\right|\frac{\partial}{\partial\varphi}\left|\begin{array}{c}\sqrt{\frac{1-\cos\theta}{2}}e^{-2i\varphi}\\\sqrt{\frac{1+\cos\theta}{2}}\end{array}\right\rangle\right) = 1 \quad (17)$$

We see that the Berry curvature is constant in these coordinates, equivalent to the Gauss curvature of a sphere. Its integral is a topological invariant and has a value of $4\pi$, which corresponds to a Chern number 2:

$$C = \frac{1}{2\pi}\int \mathbf{B}\cdot\mathbf{dS} \quad (18)$$

Large Chern numbers are a special field of research since Ref.[37]. Indeed, most of the previous studies were dealing with Chern numbers $\pm 1$ (and consequently, one edge state for each edge). Larger Chern numbers are fundamentally different, because they allow the creation of quantum superpositions of the edge states (indeed, more than 1 state is required for a superposition). From the applied point of view, more edge states means higher conduction and smaller resistance, increased mode density and coupling efficiency. With two chiral photonic edge states, one can fabricate a reflectionless waveguide splitter, which has a clear technological interest.

A similar calculation can be done for a fermion (e.g. an electron), whose pseudospin rotates on the unit sphere (Bloch sphere) under the effect of an applied magnetic field. In this case, the field makes a single rotation around the vertical axis, and therefore the Berry curvature is twice smaller: $B_r = 1/2$. The corresponding Chern number is equal to 1. For the photon in the TE-TM field it is twice larger because of the double in-plane rotation. This constant curvature is the reason why the Bloch sphere is a good representation for the pseudospin: indeed, it simply tells us that the wavefunction described by the pseudospin changes from point to point in exactly the same way as an ordinary vector, normal to the surface of a geometric sphere. The value of the curvature will be important for the calculations below. To conclude, the integral of the full sphere is $2\pi$ for single in-plane winding (e.g. Rashba, monopolar and Dresselhaus fields) and $4\pi$ for double in-plane winding (TE-TM field).

**III Photonic topological insulator**

Since the topological invariant (the Chern number) of the Bloch sphere is 1, we can be sure that a given energy band is topologically non-trivial if, when visiting adiabatically all the band states within the first Brillouin zone, the pseudo spin executes a complete wrapping of the Bloch sphere, from one pole to another. This is realized, for example, if the wavefunction can be written in the form

$$\psi_{\vec{k}} = \frac{1}{\sqrt{2}}\begin{pmatrix}\sqrt{1+\rho_c^k}e^{-i\varphi}\\\sqrt{1-\rho_c^k}e^{i\varphi}\end{pmatrix} \quad (19)$$

with $\rho_c$ going from -1 to +1 when going from the $\Gamma$ point to the edge of the Brillouin zone. If within a given energy band, there are several energy branches, the band Chern number is the sum of the Berry curvature accumulated over these branches. The wavefunction just above accumulates a Berry curvature $4\pi$ over the Brillouin zone. If we consider, however, a periodic system without any polarization splitting, the modes will all be doubly degenerate. One can perform the following decomposition:

$$\psi_{\vec{k}}^{+} = \frac{1}{\sqrt{2}}\begin{pmatrix} \sqrt{1+\rho_c^k}\,e^{-i\varphi} \\ \sqrt{1-\rho_c^k}\,e^{i\varphi} \end{pmatrix},\ \psi_{\vec{k}}^{-} = \frac{1}{\sqrt{2}}\begin{pmatrix} -\sqrt{1-\rho_c^k}\,e^{-i\varphi} \\ \sqrt{1+\rho_c^k}\,e^{i\varphi} \end{pmatrix} \quad (20)$$

These two states have a pseudospin parallel and anti-parallel to a vector $\vec{\Omega} = \vec{\Omega}_{TETM} \pm \Omega_z \vec{u}_z$, with $\Omega_{TETM}$ which should cancel at both $\Gamma$ and the Brillouin zone edge. Their in-plane parts have an opposite sign but the same winding, whereas their circularity varies in an opposite way going along the Brillouin zone. As a result, the Berry phases accumulated by two wavefunctions cancel each other.

**III-a The Haldane-Raghu proposal of a Photonic topological insulator [13]**
The Haldane-Raghu proposal was to consider a photonic crystal waveguide with a honeycomb periodicity and a splitting between TE and TM modes. Figure 9 shows the reciprocal space of a honeycomb lattice with the hexagonal Brillouin zones. The red arrows show the effective TE-TM field which is so large that we can consider only the TE state with the pseudo spin aligned with the effective field. The zoom on the corners of the Brillouin zone shows that the texture around these points is the one of a Dresselhauss field centered on K and K'. We can speak of the emergence of a local Dresselhauss gauge field [38,39,40]. These effective fields have opposite signs but the same winding. In a photonic crystal waveguide, guided modes with $k = 0$ do not exist and the field does not have to cancel there.

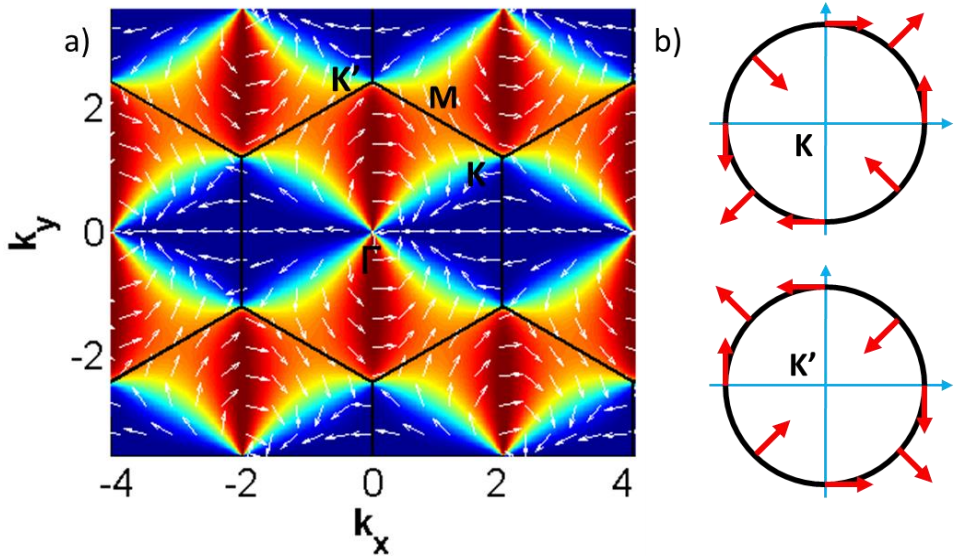

*Figure 9. a) The texture of the TE-TM field in several Brillouin zones, naturally giving rise to Dresselhaus field at the corners; b) Texture of the Dresselhaus field in K and K' points, showing opposite signs but identical windings.*

Then, on top of that, Haldane-Raghu proposed to consider a Zeeman-like (gyromagnetic) response of the photonic crystal, which in practice can have various origins [14], but limits the physical implementation to the microwave range. This Zeeman-like splitting plays a very little role everywhere in the Brillouin zone, being much smaller than the in-plane TE-TM field, except close to the Dirac point, where a Zeeman-induced gap opens. Approaching K or K', the pseudospin becomes more and more circular, while the in-plane pattern around these points is vortical. The Berry curvature accumulated around each of the Dirac points is $\pi$, because the field texture is the Dresselhaus one ($2\pi$ for the whole sphere) and only ½ of the Bloch sphere is covered, since the band polarization evolves from linear at k=0 to circular at the Dirac point. As a result, the whole valence band is topologically non-trivial, equivalent to a complete Bloch sphere, whose Chern number is 1. The conduction band has an opposite Chern number -1, because it has the same winding, but opposite integration direction over $\rho_c$. Using the analogy with the Möbius strip, there are two half-integer twists in the same direction, which add together to give a full integer twist. The paradromic rings obtained by cutting such strip into two cannot be separated (see Fig. 10a), similar to the valley and conduction bands of the topological insulator (Fig. 10b), which remain coupled by the obligatory edge state.

### III-b Polariton Z topological insulator

Let us now consider the polariton graphene, based on a honeycomb lattice of pillar microcavities [41]. The ingredients here are essentially the same as in the Haldane-Raghu proposal: a spin-orbit coupling and a time-reversal symmetry breaking. However, the regimes for both are different. In microcavities, the $\Gamma$-point exists, the dispersion close to $k=0$ is parabolic with a vanishing TE-TM splitting. The other difference with the Haldane-Raghu proposal is that the origin of the Zeeman field is the Zeeman splitting of excitons, which is non-zero for optical frequencies. When a non-zero Zeeman field is present, the k=0 states split in two circularly-polarized states, and progressively become TE and TM when the wavevector increases. Each of these two sub-bands provides an opposite contribution to the Berry curvature which therefore cancel out. Thus, the $\Gamma$-point does not contribute to the Chern number. Let us now look at the pattern close to the Dirac points. In the realistic limit of low TE-TM splitting ($\delta J < 0.5 J$, where $J$ is the mean tunneling constant and $\delta J$ is the difference between the tunneling coefficients for each polarization component), a complicated trigonal warping effect is taking place. The corresponding term mixes the two pseudospins (polarization and sublattice). The Dirac point is split in four. When a gap is opened by a Zeeman field, the central Dirac point gives a contribution to the Berry curvature which is opposite to contribution of the three others, and the total Chern number of the valence band is 2. This weak TE-TM limit [28,29] is different from the typical case of photonic crystals. The band structure of such polariton topological insulator is shown in Fig. 7b, calculated using the tight-binding model for a sample consisting of 50 zigzag chains and infinite in the perpendicular direction (marked *x*). The black lines correspond to the states in the conduction and valence bands (quantized in the transverse direction). Color shows the edge localization degree of a state, obtained from its eigenvector $\Psi$ as the relative difference between the densities on the left and right edges of the structure).

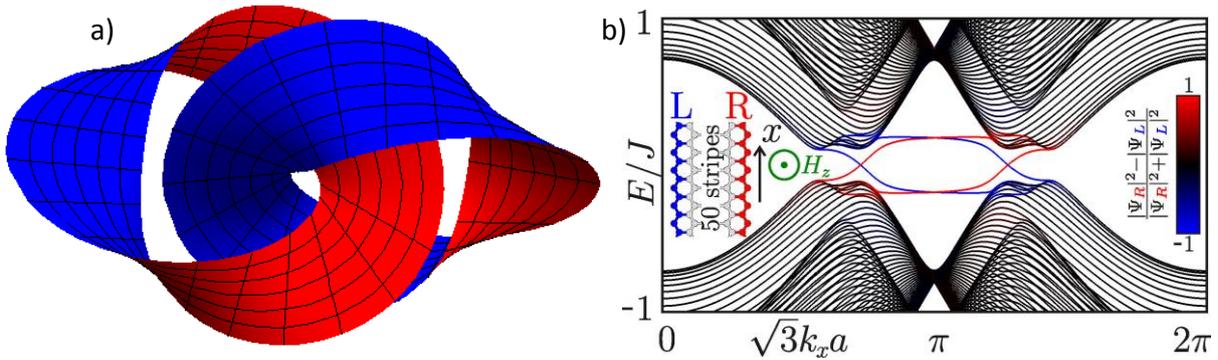

*Figure 10. a) Paradromic rings from a full-twist Möbius strip; b) Valence and conduction bands of a polariton Z topological insulator, joined by edge states [28].*

For the rest of this chapter, we consider the simpler case of the large TE-TM splitting $\delta J > 0.5 J$. The original Hamiltonian matrix from [28,39,40] in the basis $\left( A^+, A^-, B^+, B^- \right)^T$ of circular-polarized states localized on pillars A, B reads:

$$H_0 = H_D + H_{SO} = \begin{pmatrix} 0 & 0 & qe^{-i\phi} & 2\Delta \\ 0 & 0 & 0 & qe^{-i\phi} \\ qe^{i\phi} & 0 & 0 & 0 \\ 2\Delta & qe^{i\phi} & 0 & 0 \end{pmatrix} \quad (21)$$

This expression is obtained by the linearization of the graphene Hamiltonian (5) which gives the Dirac Hamiltonian for graphene $H_D = \hbar v_F \boldsymbol{\sigma}\mathbf{k}$ with spin-orbit coupling Hamiltonian $H_{SO}$ [28] ($\hbar v_F = 3Ja/2$). Here we restrain ourselves to the vicinity of the Dirac point K, which becomes the new origin of the coordinate system $q, \phi$. The $q$-linear terms appear from the original graphene Hamiltonian (5), while the $2\Delta$ term comes from the spin-orbit coupling, which is constant in the first order of $q$. The assumption of large TE-TM splitting allows to avoid dealing with the trigonal warping. To simplify the consideration, we take $\hbar v_F = 1$, and $\Delta$ determines the TE-TM splitting magnitude $\delta J$ ($\Delta = 3\delta J/2$). Complete diagonalization gives a dispersion, which is parabolic at small wavevectors and linear at large wavevectors:

$$E = \pm\Delta \pm \sqrt{\Delta^2 + q^2} \quad (22)$$

The corresponding eigenvectors in the order of increasing eigenvalues are:

$$\psi_1 = \left( -\frac{1}{2}\sqrt{1 + \frac{\Delta}{\sqrt{\Delta^2 + q^2}}}, -\frac{qe^{-i\phi}}{2\sqrt{\Delta^2 + q^2 + \Delta\sqrt{\Delta^2 + q^2}}}, \frac{qe^{i\phi}}{2\sqrt{\Delta^2 + q^2 + \Delta\sqrt{\Delta^2 + q^2}}}, \frac{1}{2}\sqrt{1 + \frac{\Delta}{\sqrt{\Delta^2 + q^2}}} \right)$$

$$\psi_2 = \left( \frac{1}{2}\sqrt{1 - \frac{\Delta}{\sqrt{\Delta^2 + q^2}}}, -\frac{qe^{-i\phi}}{2\sqrt{\Delta^2 + q^2 - \Delta\sqrt{\Delta^2 + q^2}}}, -\frac{qe^{i\phi}}{2\sqrt{\Delta^2 + q^2 - \Delta\sqrt{\Delta^2 + q^2}}}, \frac{1}{2}\sqrt{1 - \frac{\Delta}{\sqrt{\Delta^2 + q^2}}} \right)$$

$$\psi_3 = \left( -\frac{1}{2}\sqrt{1 - \frac{\Delta}{\sqrt{\Delta^2 + q^2}}}, \frac{qe^{-i\phi}}{2\sqrt{\Delta^2 + q^2 - \Delta\sqrt{\Delta^2 + q^2}}}, -\frac{qe^{i\phi}}{2\sqrt{\Delta^2 + q^2 - \Delta\sqrt{\Delta^2 + q^2}}}, \frac{1}{2}\sqrt{1 - \frac{\Delta}{\sqrt{\Delta^2 + q^2}}} \right)$$

$$\psi_4 = \left( \frac{1}{2}\sqrt{1 + \frac{\Delta}{\sqrt{\Delta^2 + q^2}}}, \frac{qe^{-i\phi}}{2\sqrt{\Delta^2 + q^2 + \Delta\sqrt{\Delta^2 + q^2}}}, \frac{qe^{i\phi}}{2\sqrt{\Delta^2 + q^2 + \Delta\sqrt{\Delta^2 + q^2}}}, \frac{1}{2}\sqrt{1 + \frac{\Delta}{\sqrt{\Delta^2 + q^2}}} \right)$$

As can be easily seen, the relative phase between the components of the polarization spinor is the same on A and B atoms. The spinor for the lowest energy state, shown in red in the above equation, is given by:

$$\begin{pmatrix} 1 \\ e^{-i\phi} \end{pmatrix} \quad (23)$$

whose rotation corresponds to that of the Dresselhaus field: $\vec{S} = (\cos\phi, -\sin\phi, 0)$. For all other states it is also either aligned or anti-aligned with the field. Therefore, as expected from Fig. 6, the Berry curvature close to the Dirac point is that of the Dresselhaus field $B_r = 1/2$.

One can also see that the circular polarization degree given by

$$\rho_{1A}^c = 1 - \frac{k^2}{2d^2}, \quad \rho_{1B}^c = -1 + \frac{k^2}{2d^2} \tag{24}$$

is opposite on A and B. The middle states (which have $E=0$ for $q=0$) have inverted circular polarization degree with respect to the split-off states ($E = \pm 2\Delta$ for $q=0$), as can be expected from the shape of the Hamiltonian exactly at the K point. At the K' point of the reciprocal space, there is an additional phase between bonding and anti-bonding pairs, but the Dresselhaus field remains unchanged, and the contributions to the Chern integral do not cancel out, contrary to the in-plane field of the Dirac Hamiltonian for graphene. The same reasoning as for the Haldane-Raghu proposal allows to conclude, that since each of the two Dirac points covers ½ of the Bloch sphere itself having a curvature ½, the total Chern number will be equal to 1 when the gap is opened by a field breaking the time-reversal symmetry.

**III-c. Topolaritons**

In Ref. [42], Karzig, Bardyn, Linder and Refael considered the polariton mode of a 2D slab. They took into account a pure TE mode and a finite excitonic Zeeman splitting and studied the mixing between the pure TE mode and a single spin-polarized exciton. In their case, the spinor was not the standard one associated with polarization, but the one composed of the exciton $\psi_{exc}^{\pm}$ and photon $\psi_{phot}^{\pm}$ projections of the wavefunction:

$$\psi_{pol}^{\pm} = \begin{pmatrix} \psi_{phot}^{\pm} \\ \psi_{exc}^{\pm} \end{pmatrix} = \frac{1}{\sqrt{2}} \begin{pmatrix} \pm e^{-i\varphi}\sqrt{1 \pm \beta_k} \\ \sqrt{1 \mp \beta_k} \end{pmatrix} \tag{25}$$

Where $\pm$ stands for the upper and lower polariton branches respectively, and

$$\beta_k = \left(\omega_k^{phot} - \omega_k^{exc}\right) / \sqrt{\left(\omega_k^{phot} - \omega_k^{exc}\right)^2 + 4g^2} \tag{26}$$

is the detuning factor, where $\omega_k^{phot}$ and $\omega_k^{exc}$ are the frequencies of the photonic and excitonic modes at the wavevector $k$, respectively, and $g$ is one half of the Rabi splitting. It approaches -1 when $k$ goes to 0 and +1 when $k$ goes to infinity. In other words, the lower polariton branch becomes strongly photonic at $k=0$ and entirely excitonic when $k$ goes to infinity, the reverse taking place for the upper polariton mode. The "Chern" number of the lower and upper polariton modes, integrating the Berry curvature for $k$ from 0 to infinity is close to $\pm 1$ respectively. The core of the vortex, the point of reciprocal space where the wavefunctions become circularly polarized, is not provided by a crossing (Dirac) point in reciprocal space, but it corresponds to the states being fully excitonic, and fully photonic. Formally this is taking place only in the limit of infinite wavevectors. In the basic idea of the authors, a topological gap should be opened between the exciton-like top of the lower polariton branch (this "top" does only exist if the exciton mass is taken zero) and the exciton-like bottom of the upper branch. In practice, the authors conclude that a topological gap can be opened if a periodic potential for excitons is created which opens a gap in the exciton-dominated part of the spectrum. In practice, this is quite challenging. In fact, it clearly appears in the light of the forthcoming works by the same authors [43], that with the very same ingredients (TE-TM splitting, Zeeman field and, at the end, a periodic potential), it is much more realistic and simple to consider the polarization spinor, a photonic honeycomb lattice with Dirac cones, and implement a polaritonic analog of the Haldane-Raghu scheme.

**IV Discussion/Conclusion**

*On the photonic $Z_2$ topological insulator*

One should mention that several proposals and realizations of photonic analogs of a $Z_2$ topological insulator have been published in the recent years [44,45]. The "spins" which are transported on the surfaces are not necessarily related to the polarization. For example, Hafezi [44] proposed to use coupled ring resonators where the role of the spin is played by the angular momentum $\pm 1$ of the states in these rings. The question which then arises [15] is about the conservation of these "spins" in presence of disorder. Indeed, the disorder in a ring can very well scatter one angular momentum state into another and the propagating state on one surface can therefore be backscattered. This argument possibly holds even for the pseudospin associated with polarization, which, according to some authors [15] would make difficult, in principle, the realization of a true photonic analog of a $Z_2$ topological insulator.

*On the polariton topological insulator*

The first way to organize dissipationless transport is superfluidity, where a coherent flow of particles is protected by interactions. The second way relies on the topological protection, which can be achieved either using a geometrically non-trivial structure, or by engineering the band structure. The remarkable properties and the flexibility of the polariton system allow to achieve both phenomena. Superfluidity which has been predicted [46] and observed [47] for polaritons became possible thanks to their strong interacting character. Topological protection has been achieved in a geometrically non-trivial structure (ring [26]) and predicted in coupled cavity lattices [28,29] and photonic crystal slabs [43]. A remarkable aspect of polaritons is that they offer an important magnetic response at optical frequencies, a magnetic response which can even be made stronger using semi-magnetic semiconductors. In principle, the combination of large band gap semi-magnetic semiconductors allowing room temperature polaritonics [48] with large Zeeman effect and with the guided geometry allowing long propagation distances without photonic losses, could allow viable topological photonics at room temperature at optical frequencies, which is the main challenge of the field [15].

*General outlook*

Topological photonics is an extremely active branch of physics, where every year brings new ideas and discoveries. Many of proposed non-trivial topological effects have already been realized, but we expect new ones to appear in the focus of attention of the large community. One of the possible directions could be linked with interactions and collective effects [49,50,51,52], which is currently only starting to develop. We are also looking forward to the applications of chiral structures, for example, for the isolation of optical devices [53], but also to reduce the sensitivity to disorder [54,55,56,57]. Thus, topological photonics promises important breakthroughs not only for fundamental, but also for applied physics.

Acknowledgements: We thank O. Bleu for critical reading of the manuscript and support.

**References**

1 W. Gerlach, O. Stern, Zeitschrift fur Physik 9, 349 (1922); ibid 9, 353 (1922).
2 M. I. D'Yakonov and V. I. Perel, JETP Lett. 13, 467 (1971).
3 H. Kamerlingh Onnes, Proc. K. Ned. Akad. Wet. 13, 1107 (1911).
4 P. Kapitza, Nature, 141, 74 (1938); J. F. Allen, and A.D. Misener, NATURE, 141, 75 (1938).
5 M.V. Berry, Proc. R. Soc. Lond. A, 392, 45 (1984).
6 S. Pantcharatnam, Proc. Indiana, Acad. Sci. A, 44, 127 (1956).
7 S.S. Chern, Annals of Mathematics. Second Series 47, 85 (1946).
8 K.V. Klitzing, G. Dorda, M. Pepper, Phys. Rev. Lett. 45, 494 (1980).
9 Y. Hatsugai, Phys. Rev. Lett. 71 3697, (1993).
10 M. Z. Hasan and C. L. Kane, Rev. Mod. Phys. 82 ,3045 (2010).
11 F.D.M. Haldane, Phys. Rev. Lett. 61, 2015 (1988).

12 C. L. Kane and E. J. Mele, Phys.Rev.Lett. 95, 146802 (2005); C. L.Kaneand E. J.Mele, Phys.Rev.Lett. 95, 226801(2005); Bernevig, B. A., T. A. Hughes, and S. C. Zhang, Science 314, 1757, (2006).
13 F. D. M. Haldane and S. Raghu, Phys. Rev. Lett. 100, 013904 (2008).
14 Z. Wang, Y. Chong, .D. Joannopoulos, M. Soljacic, Nature, 461, 772, (2009).
15 L. Lu, J.D. Joannopoulos, M. Soljacic, Nature Phot., 8, 821, (2014).
16 J. Cho, D.G. Angelakis, S. Bose, Phys. Rev. Lett. 101, 246809 (2008).
17 M.C. Rechtsman et al., Nature, 496, 196, (2013).
18 K. Fang, Z. Yu, S. Fang, Nature Photonics 6, 236 (2012).
19 J. Koch, A.A. Houck, K. Le Hur, S. M. Girvin, Phys. Rev. A 82, 043811 (2010).
20 R.O. Umucalilar, I. Carusotto, Phys. Rev. A 84, 043804 (2011).
21 M. Hafezi, J. Mod. Phys. B 28, 1441002 (2014).
22 W. Gao et al, Phys. Rev. Lett. 114, 037402 (2015).
23 V. Yannopapas, Phys. Rev. B 83, 113101 (2011).
24 K.Y. Bliokh, D. Smirnova, F. Nori, Science 348, 1448 (2015).
25 Alexey Kavokin, Guillaume Malpuech, and Mikhail Glazov, Phys. Rev. Lett. 95, 136601 (2005). C. Leyder, et al., Nature Physics 3, 628 (2007).
26 V.G. Sala et al. Phys. Rev. X ,5 , 011034 (2015).
27 I.A. Shelykh, G. Pavlovic, D.D. Solnyshkov, G. Malpuech, Phys. Rev. Lett., 102 046407, (2009).
28 A.V. Nalitov, D.D. Solnyshkov, G. Malpuech, Phys. Rev. Lett. 114, 116401, (2015).
29 C. E Bardyn, T. Karzig, G. Refael, TCH Liew, Phys. Rev. B 161413 (2015).
30 E.I. Rashba, Sov. Phys. Solid State 1, 368 (1959); E.I. Rashba and V.I. Sheka, Fiz. Tv. Tela: Collected Papers 2, 162 (1959).
31 G. Dresselhaus, Phys. Rev. 100, 580 (1955).
32 I.A. Shelykh, A.V. Kavokin, Y.G. Rubo, T.C.H. Liew, G. Malpuech, Semic. Sc.&Techn. 25, 013001, (2010).
33 G. Panzarini et al, Phys. Rev. B 59, 5082 (1999).
34 TCH Liew et al., Phys. Rev. B 80, 161303 (2009).
35 V.S. Liberman, B.Ya. Zel'dovich, Phys. Rev. A 46, 5199 (1992).
36 K.Y. Bliokh, A. Niv, V. Kleiner and E. Hasman, Nature Photonics 2, 748 (2008).
37 S.A. Skirlo, L. Lu, M. Soljacic, Phys. Rev. Lett. 113, 113904 (2014).
38 H. Terças, H. Flayac, D.D. Solnyshkov, and G. Malpuech, Phys. Rev. Lett. 112, 066402 (2014).
39 A. Nalitov, G. Malpuech, H. Terças and D. Solnyshkov, Phys. Rev. Lett. 114, 026803, (2015).
40 D. Solnyshkov et al., arXiv:1509.04009
41 T. Jacqmin et al., Phys. Rev. Lett. Phys. Rev. Lett. 112, 116402, (2014).
42 T. Karzig, C.-E. Bardyn, N. H. Lindner, G. Refael Phys. Rev. X 5, 031001 (2015).
43 K. Yi and T. Karzig, arXiv:1510.00448.
44 M.Hafezi et al., Nature Physics 7, 907, (2011).
45 A. B. Khanikaev et al. Nature Mat., 12, 233, (2013).
46 A. Kavokin, G. Malpuech, FP Laussy, Phys. Lett. A 306, 187 (2003); I. Carusotto and C. Ciuti, Phys. Rev. Lett. 93, 166401, (2004).
47 A. Amo et al, Nature Physics, 5, 805, (2009) ; A. Amo et al. Science, 332, 6034 (2011).
48 F. Li et al, Phys. Rev. Lett. 110, 196406 (2013).
49 X. Chen et al, Science 338, 1604 (2012).
50 R.O. Umucalilar, I. Carusotto, Phys. Rev. Lett. 108, 206809 (2012).
51 M. Hafezi, M.D. Lukin, J.M. Talyor, New J. Phys. 15, 063001 (2013).
52 Y. Lumer et al, Phys. Rev. Lett. 111, 243905 (2013).
53 D. Jalas et al, Nature Photonics 7, 579 (2013).
54 Y. Yang et al, Appl. Phys. Lett. 102, 231113 (2013).
55 S. Mittal et al, Phys. Rev. Lett. 113, 087403 (2014).
56 J. Koch et al, Phys. Rev. A 82, 043811 (2010).
57 A. Aspuru-Guzik et al, Nature Comm. 3, 882 (2012).